\documentclass{ptephy_v1}

\preprintnumber{XXXX-XXXX} 

\usepackage{indentfirst}
\usepackage{color}

\allowdisplaybreaks
\begin{document}

\title{Possibility of rapid neutron star cooling with the realistic equation of state}


\author{
   Akira Dohi$^{1}$\thanks{dohi@email.phys.kyushu-u.ac.jp},
	Ken'ichiro Nakazato$^2$,
   Masa-aki Hashimoto$^1$,
  Yasuhide Matsuo$^1$,
  and Tsuneo Noda$^3$
}
\affil{
  $^1$Department of Physics, Faculty of Science, Kyushu University, 744 Motooka, Nishi-ku, Fukuoka 819-0395, Japan \\
  $^2$Faculty of Arts \& Science, Kyushu University, 744 Motooka, Nishi-ku, Fukuoka 819-0395, Japan \\
  $^3$Kurume Institute of Technology, 2228-66 Kamitsumachi, Kurume, Fukuoka 830-0052, Japan 
}


\begin{abstract}%
Whether fast cooling processes occur or not is crucial for the thermal evolution of neutron stars. In particular, the threshold of the direct Urca process, which is one of the fast cooling processes, is determined by the interior proton fraction $Y_p$, or the nuclear symmetry energy. Since recent observations indicate the small radius of neutron stars, a low value is preferred for the symmetry energy. In this study, simulations of neutron star cooling are performed adopting three models for equation of state (EoS): Togashi, Shen, and LS220 EoSs. The Togashi EoS has been recently constructed with realistic nuclear potentials under finite temperature, and found to account for the small radius of neutron stars. As a result, we find that, since the direct Urca process is forbidden, the neutron star cooling is slow with use of the Togashi EoS. This is because symmetry energy of Togashi EoS is lower than those of other EoSs. Hence, in order to account for observed age and surface temperature of isolated neutron stars (INS) with use of the Togashi EoS, other fast cooling processes are needed regardless of the surface composition.
\end{abstract}

\subjectindex{E32, D41} 

\maketitle

\section{Introduction}
Neutron stars are formed as a remnant of core-collapse supernovae. Nascent neutron stars are hot and their temperature decreases with their age. Since the emissivity of neutrinos is much larger than that of photons, the neutrino emission determines the thermal evolution of neutron stars until $10^{5-6}$ yr after the birth. Since the ages of observed neutron stars are mostly younger than $10^{6-7}$ yr \cite{Vigano2013}, the observed temperature of neutron stars depends mainly on neutrino emission processes.

The thermal evolutions of neutron stars are described by cooling curves; the relation between the age $t$ and the effective temperature $T_{\mathrm{eff}}^{\infty}$ of neutron stars. So far, many theoretical models of cooling curves have been compared to the observed temperature and age of isolated neutron stars (INS)
\cite{Tsuruta1964,Tsuruta1966,Tsuruta1979,Umeda1994,Tsuruta1998,Page2004,Yakovlev2004,Shternin2011,Noda2013}. The cooling process of neutron stars depends sensitively on the state of matter, which is described by the equation of state (EoS) beyond the nuclear saturation density. One of the most important ingredients on the thermal evolution of neutron stars is whether fast cooling processes occur or not. Especially, the direct Urca (DU) process is the strongest process in the nucleon cooling processes \cite{Lattimer1991a}. The threshold of DU process is determined by the proton fraction $Y_p$, which depends on EoS. Therefore, the EoS dependence of the cooling curves has been studied actively \cite{Lim2017,Negreiros2018}. Another significant effect on the neutron star cooling comes from the superfluidity \cite{Yakovlev2001,Page2004,Page2009,Ho2015}.  While the superfluid critical temperature is a crucial parameter to govern the efficiency of the cooling, it is uncertain especially for the triplet gap (e.g. \cite{Amundsen1985,Baldo1992,Yakovlev1999,Takatsuka2004}). Due to the superfluidity, thermodynamical quantities are affected severely through the neutrino loss rates. Therefore, to calculate cooling curves we adopt models of EoS, neutrino loss rates, and superfluid critical temperature, as indispensable physical quantities. 

In this paper, we focus on the EoS dependence of the cooling curves. Recently, several constraints on the EoS are considered. For example, the discovery of neutron stars with $\simeq 2~M_{\odot}$ \cite{Demorest2010,Antoniadis2013} provides a strong constraint because the EoSs with the maximum mass of neutron stars with $<$2$~M_{\odot}$ are inconsistent with them
(see also Ref. \cite{Cromartie2019}). Another examples are from the analysis of low mass X-ray binaries and gravitational wave from neutron star merger (GW170817) \cite{Steiner2010,Abbott2018}. They provide constraints on the mass--radius relation of neutron stars and the neutron star radius is indicated to be less than 13~km. In our computations of neutron star cooling, we adopt three models of EoS with different radius of neutron stars. Then, for the EoS model preferred by the observations of neutron star radius, we investigate whether the resultant cooling curves are consistent with the temperature observations of INS. Furthermore, we examine the possibility of rapid cooling process with different surface compositions.

Contents in this paper are as follows: In section 2, we explain basic equations of our cooling simulations, and neutrino processes related to superfluid models. In section 3, we introduce the models of EoS adopted in our computations. In section 4, we show the results of cooling simulation comparing with the observations of INS, and examine the EoS dependence and the effect of superfluidity. In section 5, we discuss the relation between the threshold of DU process and symmetry energy and examine the influence of surface composition. In section 6, we give a concluding remark and present future investigations.

\section{Setup of cooling simulation}
\subsection{Basic equations}
When nuclear burning does not occur, fundamental equations of stellar structure are described as follows \cite{Throne1977},
\begin{eqnarray}
  \frac{\partial M_{\mathrm{tr}}}{\partial r} \hspace*{-2mm}& = &\hspace*{-2mm} 4\pi r^{2} \rho~, \label{eq:1} \\
  \frac{\partial P}{\partial r}\hspace*{-2mm} & = &\hspace*{-2mm} -\frac{G\rho}{r^{2}}
      \left(1+\frac{P}{\rho c^{2}}\right)
      \left(M_{\mathrm{tr}}+\frac{4\pi r^{3}P}{c^{2}}\right) \left(1-\frac{2GM_{\mathrm{tr}}}{c^{2}r}\right)^{-1}~, \label{eq:2} \\
\frac{\partial (L_{r}e^{2\phi/c^{2}})}{\partial M_{r}}\hspace*{-2mm} & =\hspace*{-2mm} &
      -e^{2\phi/c^{2}}\left(\varepsilon_{\nu} + e^{-\phi/c^{2}}C_V\frac{\partial T}{\partial t}
      \right)~, \label{eq:3} \\
  \frac{\partial \ln T}{\partial \ln P}\hspace*{-2mm} & =\hspace*{-2mm} & \nabla_{\rm rad}~, \label{eq:4} \\
  \frac{\partial M_\mathrm{tr}}{\partial M_{r}}\hspace*{-2mm} & =\hspace*{-2mm} & \frac{\rho}{\rho_r}
  \left(1-\frac{2GM_\mathrm{tr}}{c^{2}r}\right)^{1/2}~, \label{eq:5}\\
  \frac{\partial \phi}{\partial M_\mathrm{tr}}\hspace*{-2mm} & =\hspace*{-2mm} & \frac{G(M_{\mathrm{tr}}+4\pi r^{3}P/c^{2})}
    {4\pi r^{4}\rho}\left(1-\frac{2GM_\mathrm{tr}}{c^{2}r}\right)^{-1}, \label{eq:6}  
\end{eqnarray}
with the gravitational constant $G$ and velocity of light $c$. Here, $\rho$ is the mass energy density, $\rho_r$ is the rest mass density, $P$ is the pressure, $M_\mathrm{tr}$ is the gravitational mass, $M_r$ is the rest mass inside the radius $r$, $T$ is the local temperature, $\varepsilon_\nu$ is the energy loss rate by neutrino emission, $C_V$ is the specific heat per baryon, $\nabla_{\rm rad}$ is the radiative gradient, and $\phi$ is the gravitational potential in unit mass.

From the latest neutron star mass measurements \cite{Alsing2018}, observed minimum and maximum masses are $1.174 \pm 0.004~M_{\odot}$ \cite{Martinez2015} and $2.01 \pm 0.04~M_{\odot}$
\cite{Antoniadis2013}, respectively. This range of the masses observed is consistent with simulations of supernova explosions \cite{Suwa2018}. In this study, we perform the cooling simulations of neutron stars with the masses from $1.1~M_{\odot}$ to $2.1~M_{\odot}$. For the age and the surface temperature of INS, we adopt observational data in Ref. \cite{Lim2017} for 18 INS and Ref. \cite{Heinke2010} for Cassiopeia A. Note that, while the temperature variation is observed by $Chandra$ for Cassiopeia A \cite{Wijingaarden2019} (but see also Ref. \cite{Elshamouty2013} ), we do not intend to account for it in this paper. Although the surface composition affects the photon emission rate due to the differences in the opacity $\kappa$ \cite{Noda2006,Page2006}, the surface compositions are fixed to be light elements, which are 73\% of ${}^{1}{\mathrm{H}}$, 25\% of ${}^{4}{\mathrm{He}}$, and 2\% of ${}^{56}{\mathrm{Ni}}$. The mass in the envelope is supposed to be $5.0\times 10^{-13}~M_{\odot}$, ignoring the accretion. In Section 5.2, however, we vary the surface composition and consider the pure Ni case. Furthermore, $\nabla_{\rm rad}$ depends on mainly $\kappa$, and is calculated by using public conductivity codes~\cite{Baiko2001,Potekhin2015}. We adopt the evolution code of a spherically symmetric neutron star \cite{Fujimoto1984,Matsuo2018}. As a result, we construct cooling curves by solving the structure equations Eqs. (\ref{eq:1}) -- (\ref{eq:6}) from the core to the surface of neutron stars. 

\subsection{Neutrino emission processes}

So as to evaluate the neutrino energy loss rate $\varepsilon_\nu$ in Eq. (\ref{eq:3}), we adopt the neutrino emission processes listed in Table 1. The modified Urca and the nucleon pair bremsstrahlung processes always turn on and their loss rate is approximately $10^{19-21} ~\mathrm{erg~cm^{-3}~s^{-1}}$ \cite{Shapiro2008}. While they are slow cooling processes, the DU process is fast cooling process. The emissivity is around $10^{27}~\mathrm{erg~cm^{-3}~s^{-1}}$ \cite{Lattimer1991a}. Since the DU process occurs satisfying the momentum conservation, the threshold of proton fraction $Y_p^{\mathrm{DU}}$ for the DU process to turn on is written as \cite{Lattimer1991a}:
\begin{equation}
Y_p^{\mathrm{DU}} = \frac{1}{1 + \left(1 + x_e^{1/3}\right)^3}, \label{eq:7}
\end{equation}
where $x_e = Y_e/(Y_e + Y_{\mu})$, defining $Y_e$ and $Y_{\mu}$ as electron and muon fractions, respectively. In the very high-density regions, $Y_p^{\mathrm{DU}} \simeq 0.1477$ because of $Y_e \simeq Y_{\mu}$. When muons are absent, $Y_p^{\mathrm{DU}}$ is $1/9$. Since $Y_p$ increases with density in high-density regions, the DU process occurs generally inside massive neutron stars.

\begin{table}
\caption{
Nucleon neutrino emission processes adopted in this study. $l$ means electron or muon.
}
\label{Tab:Nu}
\hspace*{0.5cm}
\begin{tabular}{llc} 
\hline\hline 
     Name           &\hspace*{3.0cm}Process& \hspace*{-.2cm}Efficiency  \\ 
\hline 
\vspace{3pt}
\parbox[c]{3cm}{Modified Urca\\(neutron branch)} &
$\hspace*{1cm}\begin{cases} n+n^\prime \rightarrow p+n^\prime+l^-+\bar\nu_l \\ p+n^\prime+l^- \rightarrow n+n^\prime+\nu_l \end{cases}$  & Slow 
\\
\vspace{3pt}
\parbox[c]{3cm}{Modified Urca\\(proton branch)}  &
$\hspace*{1cm}\begin{cases} n+p^\prime \rightarrow p+p^\prime+l^-+\bar\nu_l \\ p+p^\prime+l^- \rightarrow n+p^\prime+\nu_l \end{cases}$  & Slow \\
\vspace{3pt}
Bremsstrahlung          &
$\hspace*{1cm}
\begin{cases} 
n+n^\prime \rightarrow n+n^\prime+\nu_{l}+\bar\nu_{l} \\ 
n+p \rightarrow n+p+\nu_{l}+\bar\nu_{l}  \\
p+p^\prime \rightarrow p+p^\prime+\nu_{l}+\bar\nu_{l}\\
\end{cases}
\hspace*{-1cm}$ & Slow \\
\vspace{3pt}
\parbox[c]{3.5cm}{Nucleon pair \\ breaking formation \\ (PBF)}          &
$\hspace*{1cm}
\begin{cases}    
n+n \rightarrow [nn] +\nu_{l}+\bar\nu_{l} \\ 
p+p \rightarrow [pp] +\nu_{l}+\bar\nu_{l} 
\end{cases}$  
& Medium  (if $T \sim T_{\mathrm{cr}}$)\\
\vspace{3pt}
\parbox[c]{3.5cm}{Nucleon direct Urca \\ (DU)} &
$\hspace*{1cm}\begin{cases} n \rightarrow p+l^-+\bar\nu_l \\ p+l^- \rightarrow n+\nu_l \end{cases}$              
& Fast (if $Y_p \ge Y_p^{\mathrm{DU}}$)\\
\hline
\end{tabular}

\end{table}
\subsection{Nucleon superfluidity}
Nucleon superfluid effect is also important for neutron star cooling. When the local temperature $T$ is lower than the critical temperature $T_{\mathrm{cr}}$, proton--proton pairs and neutron--neutron pairs are formed. Then, the neutrinos and antineutrinos are produced, which is called the nucleon pair breaking formation (PBF) process \cite{Page2004,Flowers1976,Voskresensky1987}.  As a consequence both protons and neutrons become ${}^1S_0$ states for low-density regions and neutrons become ${}^3P_2$ state for high-density regions. As a result, neutrino emissivity is suppressed in proportional to $\exp\left(-aT_{\mathrm{cr}}/T\right)$, where $a$ is 1.76 for ${}^1S_0$ state and 8.40 for ${}^3P_2$ state \cite{Takatsuka1972}. However, the relation between the superfluid energy gap  ($\Delta=aT_{\mathrm{cr}}$) and the critical temperature is rather uncertain because of the uncertainty of the nuclear force. While many studies about nucleon superfluidity have been done
\cite{Tsuruta1998,Takatsuka1993,Takatsuka2004}, $T_{\mathrm{cr}}$ dependence on the Fermi wave number is still uncertain. We adopt CLS model as the superfluidity of ${}^1S_0$ neutrons, and four models as the other superfluidities: Combination of two ${}^1S_0$ proton superfluid models (AO \& CCDK) and two ${}^3P_2$ neutron superfluid models (BEEHS \& EEHO) \cite{Ho2015} as shown in Fig.1.
\begin{figure}[t]
\centering
\includegraphics[width=5.0in]{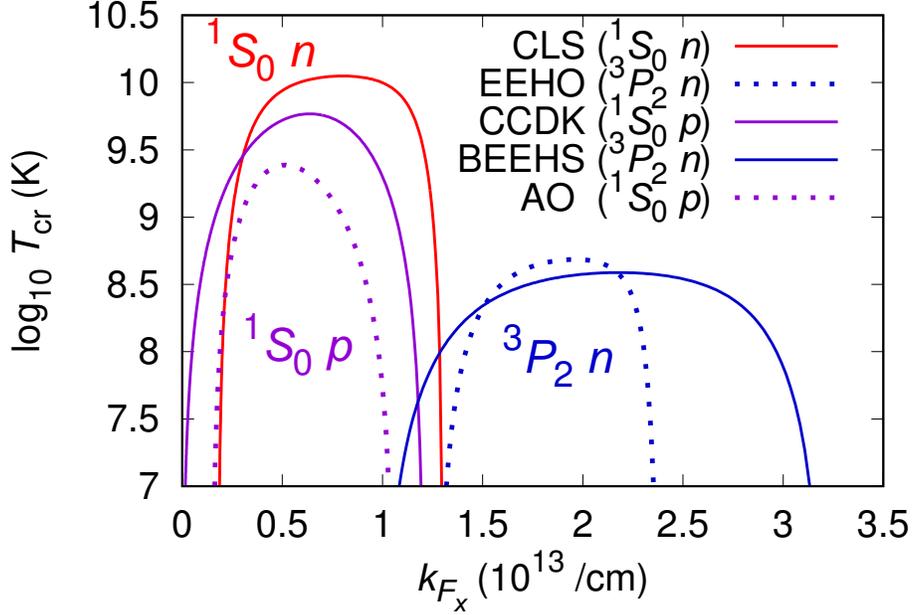}
\caption{Superfluid transition temperature vs. the Fermi wave number of a particle $x = n \ \text{or} \ p$
\cite{Ho2015}. Abbreviations of superfluid models are inserted inside the panel.}  
\end{figure}

\section{Equation of state}\label{cs}
As already noted, the EoS is important to calculate the thermal evolution of neutron stars. In this study, so as to include the temperature-dependent structure of neutron stars, we adopt three models of EoS constructed under finite temperatures: Shen, LS220, and Togashi EoSs. While the Shen EoS is based on a relativistic mean field model \cite{Shen1998a,Shen1998b,Shen2011}, the LS220 EoS is based on a Skyrme energy-density functional \cite{Lattimer1991b}. These EoSs have been widely used to simulate various astrophysical phenomena
\cite{Couch2013,Suwa2013,Famiano2016,Kuroda2017,Sumiyoshi1995,Camelio2017,Nakazato2018,Nakazato2019}. The Togashi EoS \cite{Togashi2017} is constructed with use of realistic two-body potential and phenomenological three-body potential \cite{Wiringa1995,Carlson1983,Pudliner1995} under the finite temperature. In this section, we describe the properties of these EoSs. Note that, in our calculations, while these EoSs are adopted for the pressure $P > 10^{30}~\mathrm{dyne}~\mathrm{cm}^{-2}$, we adopt BPS EoS \cite{Baym1971} in the low-density regions.

\begin{figure}[t]
\begin{minipage}{0.50\hsize}
\centering
\includegraphics[width=3.5in,height=2.7in]{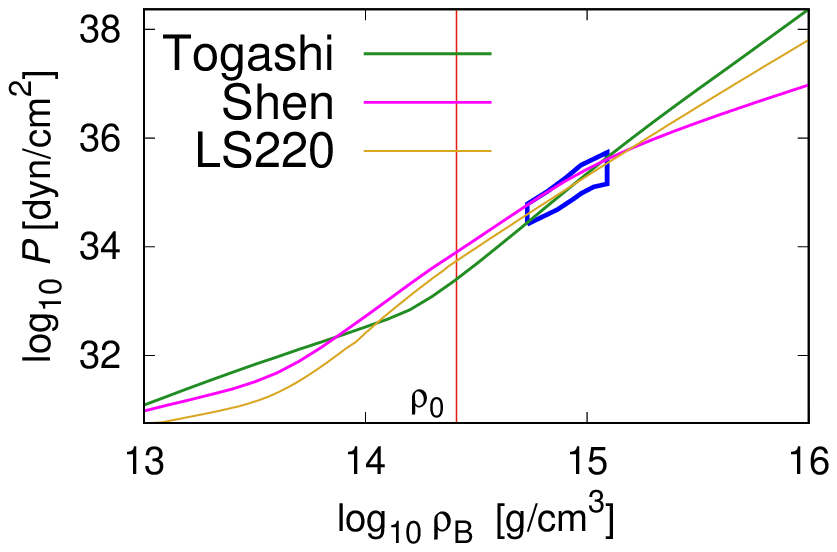}
\end{minipage}
\begin{minipage}{0.50\hsize}
\centering
\includegraphics[width=3.5in]{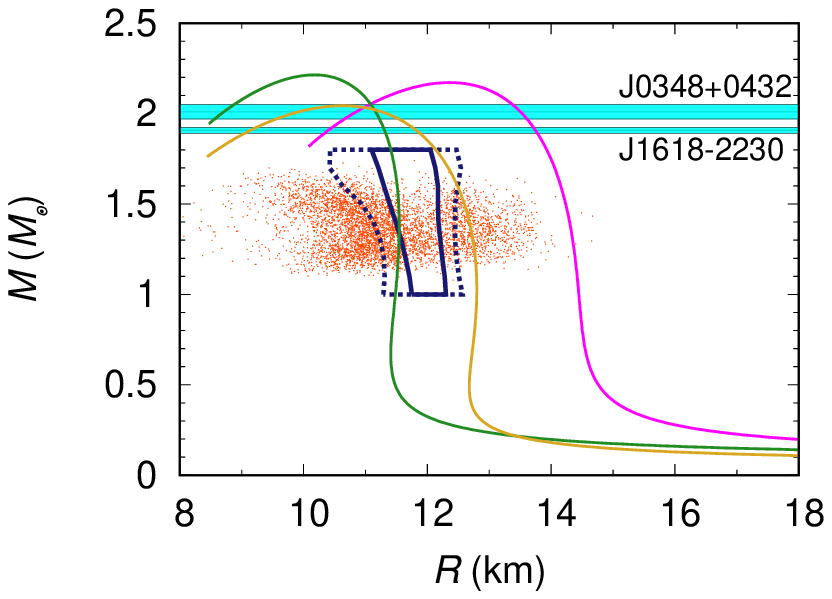}
\end{minipage}
\caption{Properties for three nuclear EoSs: Togashi (green), Shen (pink), and LS220 (orange). Left panel: Relation between baryon density and total pressure. The red vertical line indicates nuclear saturation density $\rho_0$ of the LS220 EoS. Blue region is from the flow in experiment of heavy ion collisions
\cite{Danielewicz2002}. Right panel: Mass--radius relation of neutron stars. Light blue bands show two measurements of $1.908 \pm 0.016~M_{\odot}$ of pulsar J0348+0432 \cite{Demorest2010} and $2.01 \pm 0.04~M_{\odot}$ of pulsar J1618-2230 \cite{Antoniadis2013}. Many dark orange dots indicate the results from the observation of GW170817 \cite{Abbott2018}, and blue regions indicate the results based on the low mass X-ray binary observations \cite{Steiner2010} (Solid curve: 1$\sigma$, Dashed curve: 2$\sigma$).}
\label{f2}
\end{figure}

In the left panel of Fig.~\ref{f2}, we show total pressures against the baryon density for the EoS models adopted in this study. We find that the Togashi EoS is softer than the other two EoSs for $\rho \lesssim 10^{15}$ g~$\mathrm{cm}^{-3}$. A constraint obtained from a flow measurement in experiments of heavy ion collisions \cite{Danielewicz2002} is also shown in the left panel of Fig.~\ref{f2} and we can recognize that the EoSs are consistent with the experimental results of heavy ion collisions. 

Defining $u = \rho_{\mathrm{B}}/\rho_0$ with the baryon mass density $\rho_{\mathrm{B}}$, the energy per nucleon $w(u,Y_p)$ is approximately expanded around the saturation density \cite{Vidana2009},
\begin{eqnarray}
w(u,Y_p) = w_0 + \frac{K}{18}\left(u - 1\right)^2 + \cdots + \left[S_0 + \frac{L}{3}\left(u - 1\right) + \cdots \right]\left(1 - 2Y_p \right)^2, \label{eq:9}
\end{eqnarray}
where $w_0$, $K$, $S_0$, and $L$ are, respectively, the energy per nucleon, incompressibility, symmetry energy, and symmetry energy slope at the nuclear saturation density $\rho_0$. The term inside the square bracket of Eq. (\ref{eq:9}) is a density-dependent symmetry energy $S(u) = S_0 + \frac{L}{3}\left(u - 1\right) + \cdots$. In the vicinity of the saturation density, the properties of nuclear matter are characterized by the saturation parameters, $\rho_0$, $w_0$, $K_0$, $S_0$, and $L$, and they are tabulated in Table 2 for the EoSs adopted in this study. Recent terrestrial experiments suggest $220~ \mathrm{MeV} \lesssim K \lesssim 260~\mathrm{MeV}$, $S_0 \lesssim 36~\mathrm{MeV}$, and $L \lesssim 80~\mathrm{MeV}$ \cite{Garg2018,Lattimer2013b}. Therefore, the Togashi and LS220 EoSs are consistent with the experimental values while the Shen EoS has large values of $S_0$ and $L$.
\begin{table}[t]
\begin{center}
\caption{Physical quantities at the nuclear saturation density for three EoSs}\vspace*{-0cm}
\scalebox{1.0}{
\begin{tabular}{cccccc}
\hline\hline
EoS & $\rho_0$ [$10^{14}$~g~$\mathrm{cm}^{-3}$] & $w_0$~[MeV] & $K$~[MeV] & $S_0$~[MeV] & $L$~[MeV]\\
\hline
Togashi & $2.66$ & $-16.0$  & $245$ & $30.0$ & $35.0$ \\
Shen & $2.41$ & $-16.3$  & $281$ & $36.9$ & $111$ \\
LS220 & $2.57$ & $-16.0$  & $220$ & $28.6$ & $73.8$ \\
\hline
   \end{tabular}
}
\end{center}
\end{table}

The mass--radius relations of neutron stars are plotted in the right panel of Fig.~\ref{f2} for the EoSs adopted in this study. In the present paper, we ignore the effects of rotation and magnetic field of neutron stars. Therefore, provided the relation between the total mass energy density $\rho$ and the pressure $P$, the gravitational mass $M$ and the radius $R$ of neutron stars are obtained from TOV equation as seen from Eqs. (\ref{eq:1}) and (\ref{eq:2}) \cite{TOV1939}. In the right panel of Fig.~\ref{f2}, we also show the suggestions from the observations: masses of heavy neutron stars observed so far \cite{Demorest2010,Antoniadis2013}, gravitational wave GW170817 emitted from neutron star merger \cite{Abbott2018}, and flux from low mass X-ray binaries \cite{Steiner2010}. The maximum mass of neutron stars is $2.21~M_{\odot}$ for the Togashi EoS, $2.17~M_{\odot}$ for the Shen EoS, and $2.04~M_{\odot}$ for the LS220 EoS. Hence, the three EoSs are consistent with the mass measurements of the heavy neutron stars. On the other hand, these EoSs are different with respect to the radius of neutron stars. The observations of the neutron star merger and the low mass X-ray binaries suggest the smaller radius of neutron stars less than around $13$ km. Thus, the Togashi EoS is the most preferred model among the EoSs adopted in this study. Note that, the Togashi EoS has the smallest value for $L$ and it is consistent with the fact that $R$ and $L$ have a positive correlation (e.g. $R \propto L^{1/4}$ \cite{Lattimer2001,Lim2017}) in general. 

\section{Results}
\begin{figure}[t]

\centering
\includegraphics[width=\linewidth]{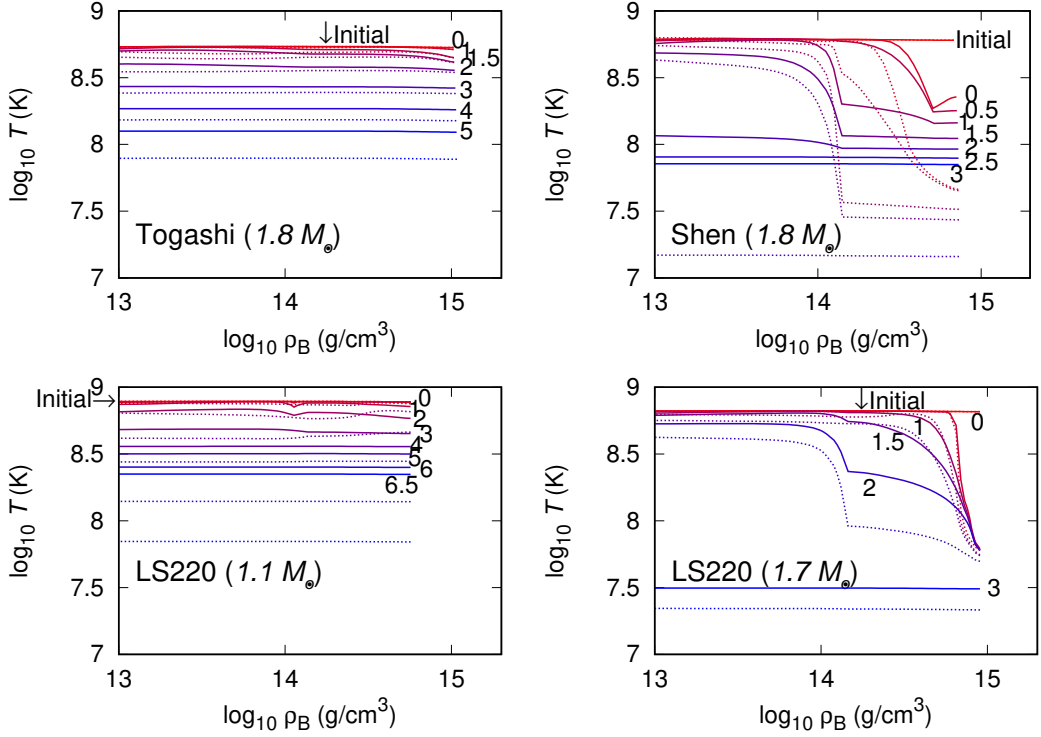}
\caption{Local temperatures against the baryon density for three EoSs (Upper left panel: Togashi with $1.8~M_{\odot}$. Upper right panel: Shen with $1.8~M_{\odot}$. Lower left panel: LS220 with $1.1~M_{\odot}$. Lower right panel: LS220 with $1.7~M_{\odot}$). Dotted curves indicate the cases without superfluid effect on neutrino emissions. Solid curves indicate the cases with superfluid models of CCDK for ${}^1S_0$ proton state and EEHO for ${}^3P_2$ neutron state. The numerals attached to the curves show the ages of log $t$ (yr).}
\end{figure}
\begin{figure}[t]
\centering
\includegraphics[width=5.0in]{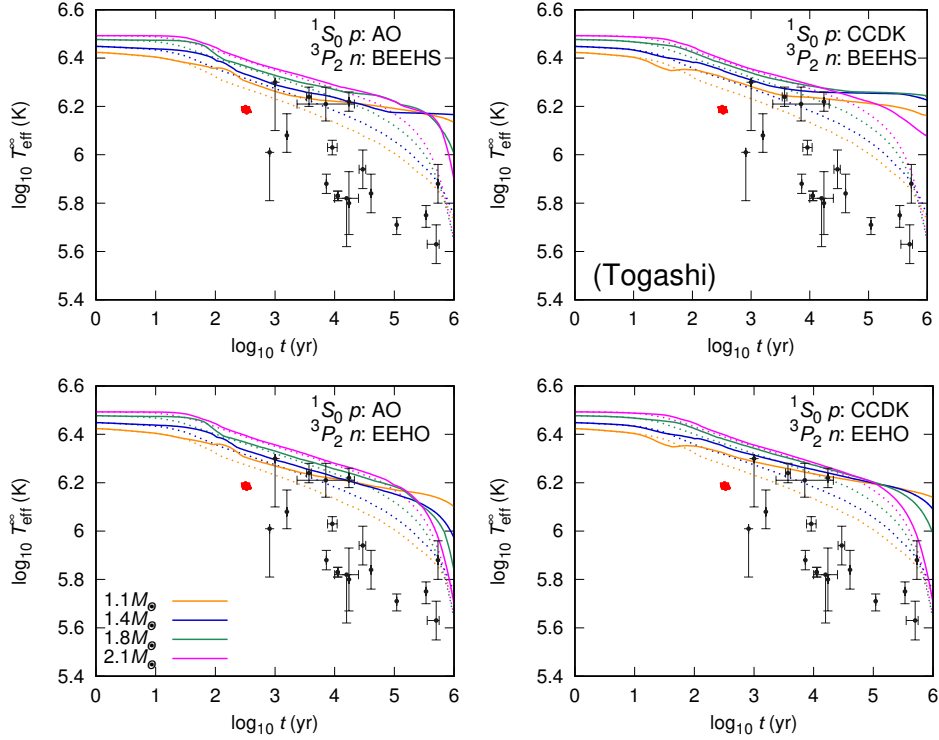}
\caption{Cooling curves using the Togashi EoS. Difference of color indicates that of gravitational mass. Dotted curves indicate the cases without the superfluidity, while solid curves are obtained by using superfluid model. The superfluid models are as follows; The left top panel: AO for ${}^1S_0$ protons and BEEHS for ${}^3P_2$ neutrons, the right top panel: CCDK for ${}^1S_0$ protons and BEEHS for ${}^3P_2$ neutrons, the left bottom: AO for ${}^1S_0$ protons and EEHO for ${}^3P_2$ neutrons, the right bottom panel: CCDK for ${}^1S_0$ protons and EEHO for ${}^3P_2$ neutrons. For the cooling data of Cassiopeia A (red symbols), we adopt the data in Ref. \cite{Heinke2010}. The others are taken from Ref. \cite{Lim2017} (black dots with errors). }
\end{figure}
\begin{figure}[htbp]
\begin{minipage}{1.0\hsize}
\centering
\includegraphics[width=5.0in]{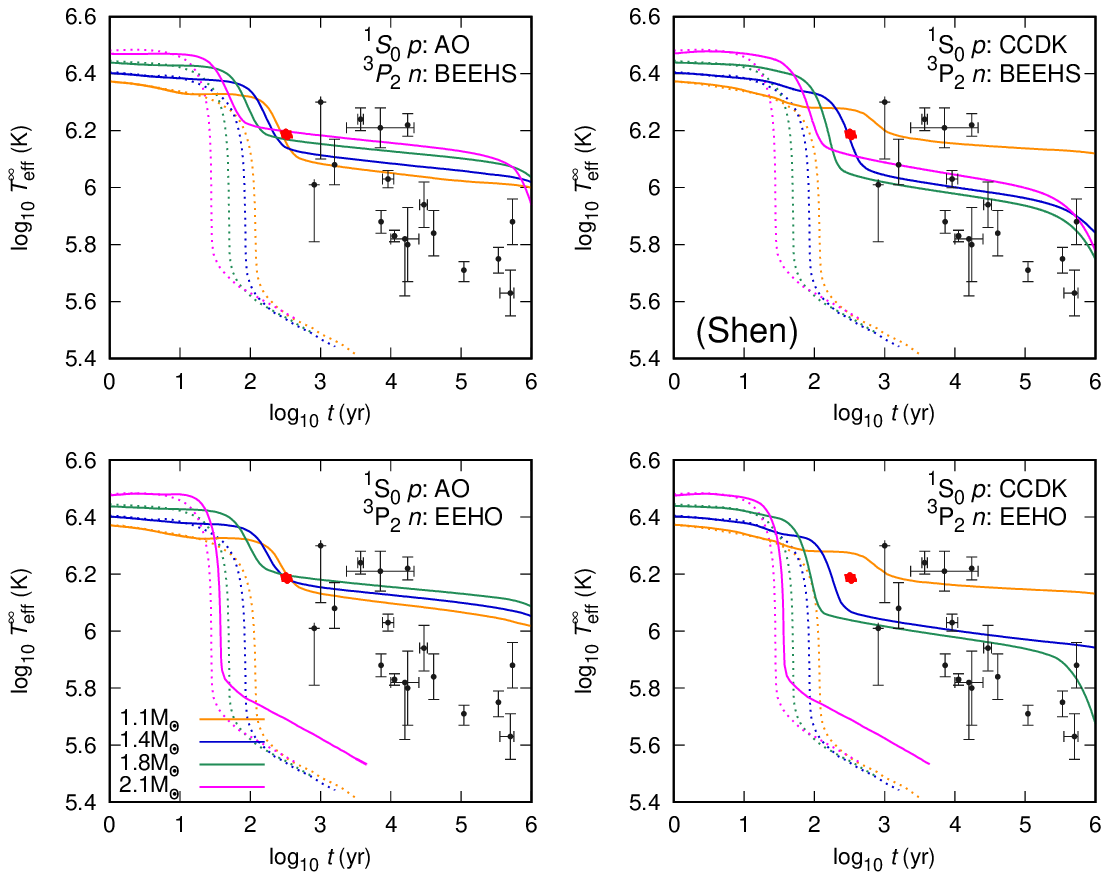}
\caption{Same as Fig. 4 but for Shen EoS}
\end{minipage}
\begin{minipage}{1.0\hsize}
\centering
\includegraphics[width=5.0in]{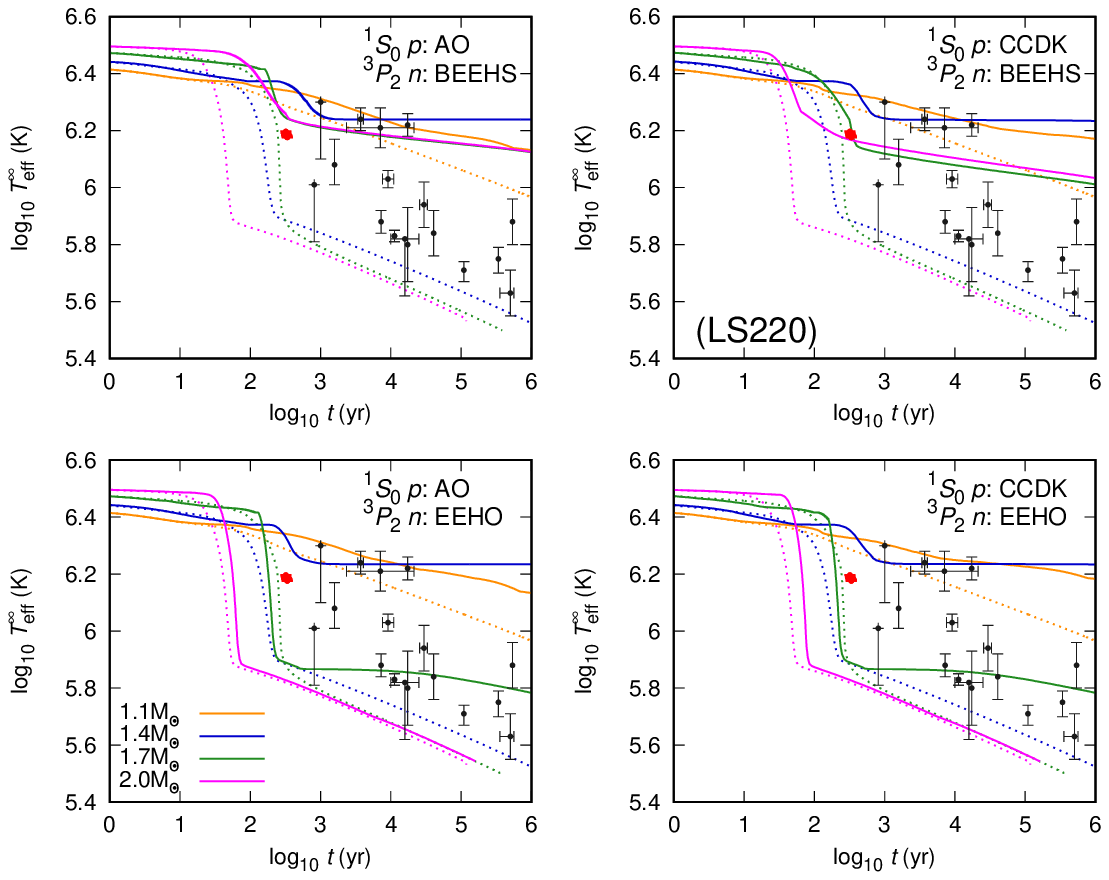}
\caption{Same as Fig. 4 but for LS220 EoS}
\end{minipage}
\end{figure}
We describe our numerical results of the neutron star cooling. The temperature profiles as a function of the baryon density are shown in Fig. 3. We adopt isothermal models for the initial conditions of the calculation except for the outer layer where $\rho_\mathrm{B}\lesssim~10^5~\mathrm{g~cm^{-3}}$. Without superfluidity, the thermal structure of the models with the Shen EoS ($1.8M_\odot$) and LS220 EoS ($1.7M_\odot$) show that the DU process works in the core of the star, and the temperatures of the cores drop rapidly. The others show that the DU process does not work and keep the thermal structure almost isothermal. It represents the slow cooling.

We present the cooling curves of neutron stars obtained from our calculations with the Togashi EoS, Shen EoS, and LS220 EoS in Figs. 4, 5, and 6, respectively. We focus on the dotted curves corresponding the cooling curves without superfluidity. In Fig. 4, the cooling curves of the Togashi EoS locate at high-temperature regions due to the slow cooling as shown in Fig. 3. This model does not account for the observational data below the curves. In Fig. 5, the cooling curves show that the models with the Shen EoS cools rapidly at any masses, and all of them do not cross observational data of INS. In Fig. 6, the cooling curves are shown for the models with the LS220 EoS. The heavy neutron stars ($M \geq 1.4 M_\odot$) cool rapidly, while the light neutron stars ($M < 1.4 M_\odot$) keep warm. By varying the mass for the models with the LS220 EoS, the cooling curves cover the range of the temperature observations. Therefore, the models with the LS220 EoS are suitable to account for the observational data.

We examine the models with superfluid effect. First, we notice the models with the Shen EoS, whose cooling curves locate lower-temperature region than all observational data without considering superfluid effect. We show the cooling curves of the models with the Shen EoS with (without) superfluid effect in the solid (dotted) curves in Fig. 5. The effective temperature for the models with the Shen EoS with superfluidity is higher than that without superfluidity, and cooling curves with superfluidity locate in higher-temperature region but are insufficient for the observations with any superfluid models. For instance, any cooling curves with the Shen EoS are inconsistent with the two observations with $t\lesssim~10^4~\mathrm{yr}$ and $T^{\infty}_{\mathrm{eff}}\gtrsim~10^{6.2}~\mathrm{K}$. Next, we focus on the models with the LS220 EoS, whose cooling curves match to the observational data without considering the superfluid effect as shown in Fig. 6. The cooling curves are sensitive to the difference of the superfluid models. If the superfluid model for $^3P_2$ neutrons is the EEHO, the cooling curves are consistent with the observations. In contrast, the models with BEEHS become inconsistent due to the quantitative difference of the superfluid effect. At high-density region, the neutron star with the LS220 EoS causes the strong DU process, which is greatly suppressed by nucleon superfluidity. Therefore, with the DU process, the superfluid effect on the cooling curves is large and the cooling becomes slow.
\begin{figure}[t]
\begin{minipage}{0.50\hsize}
\centering\hspace*{-1cm}
\includegraphics[width=3.5in]{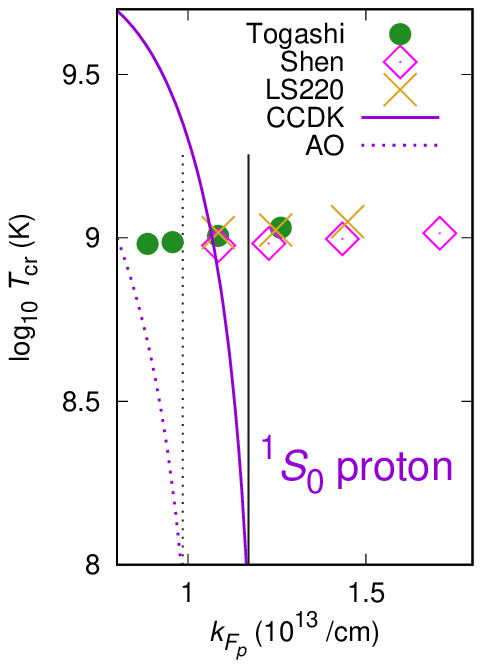}
\end{minipage}
\begin{minipage}{0.50\hsize}
\centering
\includegraphics[width = 3.5in]{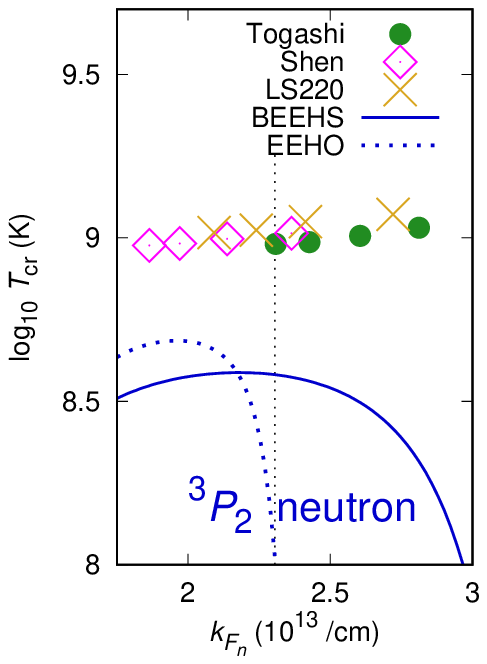}
\end{minipage}
\caption{Superfluid models vs. the central nucleon Fermi wave number for each EoS at the initial temperature. The difference in these symbols indicates that of EoS. Left and Right panels are the superfluidity for ${}^1S_0$ protons and ${}^3P_2$ neutrons, respectively. The masses corresponding to the central value of $k_{F_p}$ and $k_{F_n}$ are arranged from the left to the right $1.1, 1.4, 1.8,$ and $2.1~M_{\odot}$ for Togashi and Shen EoS, and  $1.1, 1.4, 1.7,$ and $2.0~M_{\odot}$ for LS220 EoS. These threshold values of each Fermi wave are set in case of $T_{\mathrm{cr}} = 10^8$~K.}
\end{figure}
However, without the DU process, the effect from the suppression of neutrino emission is clearly weak because the PBF process works. The PBF process is about 2--3 order of magnitude weaker than the DU process, and 1--2 order stronger than the slow cooling processes. The EoS models which do not induce the DU process such as the Togashi EoS are therefore insusceptible from the superfluidity as seen in Fig. 4. The cooling curves of the Togashi EoS locate in higher-temperature regions with considering superfluid effect, but compared to those of the Shen and LS220 EoSs, the effect of superfluidity is clearly ineffective. With and without the superfluidity, the Togashi EoS does not suit the observations for $t>10^2$ yr.

We also examine the superfluid-model dependence of the cooling curves. Without the DU process, the superfluid contribution to cooling is small as we noted. If the DU process occurs, the suppression of the neutrino emission by superfluidity is important for the neutron star cooling. In Fig. 7, the central temperatures of each model are plotted with $T_{\mathrm{cr}}$ as a function of the Fermi wave number. From this figure, we can see which model has the superfluid effect as a result of the cooling (below $10^8$ K).
For instance, considering the superfluid effect of $^1S_0$ protons for the Shen and LS220 EoSs, the threshold mass against the CCDK is about $1.1M_\odot$ and that against the AO is less than $1.1M_\odot$. This implies that the suppression of neutrino emission by the CCDK is stronger than that by AO. Figures 5 and 6 indicate such difference, but the difference is smaller than that of $^3P_2$ neutrons. As seen in Fig. 7, the threshold mass for $^3P_2$ neutrons against the EEHO for the Shen EoS is about $2.1M_\odot$ and that for the LS220 EoS is around $1.6M_\odot$. Therefore the difference of the cooling curves between $1.4M_\odot$ to $1.7M_\odot$ ($1.8M_\odot$ to $2.1M_\odot$) for the LS220 (Shen) EoS are large as shown in Fig. 6 (Fig. 5). Similarly, the cooling is slow for BEEHS as in Figs. 5 and 6 since the superfluid range of the BEEHS is wide enough to easily exceed the central Fermi wave number with $2.1M_\odot$ for any EoSs. We recognize that the best models in comparison with the INS observations are with the LS220 EoS and the EEHO $^3P_2$ neutron superfluidity.

\section{Discussion}
\subsection{The DU process and the symmetry energy at high densities}

\begin{figure}[t]
\centering
\includegraphics[width=5.0in]{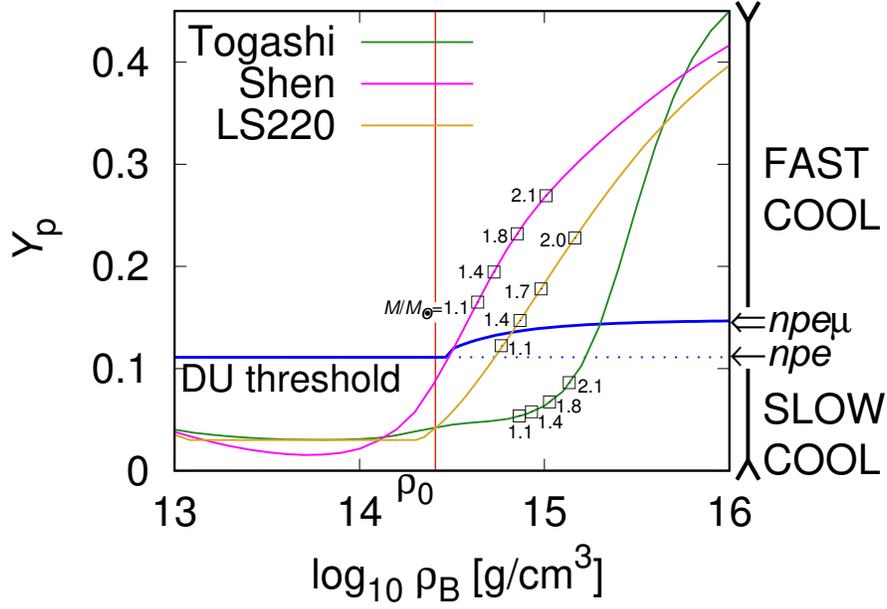}
\caption{Proton fraction $Y_p$ vs. baryon density $\rho_B$ for three nuclear EoSs: Togashi (green), Shen (pink), and LS220 (orange). Red vertical line indicates the nuclear saturation density $\rho_0$ of the LS220 EoS. Dotted blue line indicates the DU threshold in $npe$ matter while blue curve corresponds to the DU threshold in $npe{\mu}$ matter of the LS220 EoS. As in Fig. 7, $Y_p$ at the central baryon density are plotted for $1.1, 1.4, 1.8,$ and $2.1~M_{\odot}$ models with the Togashi and Shen EoSs, and  $1.1, 1.4, 1.7,$ and $2.0~M_{\odot}$ models with the LS220 EoS.}
\label{fig:yp}
\end{figure}

\begin{figure}[t]
\centering
\includegraphics[width=5.0in]{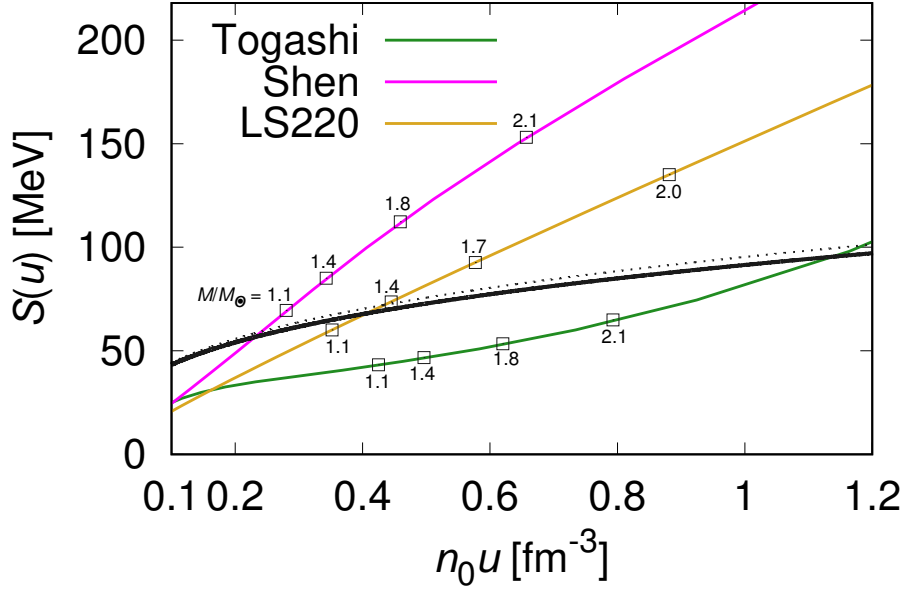}
\caption{Symmetry energies of three EoSs: Togashi (green), Shen (pink), and LS220 (orange). We adopt the saturation number density $n_0 = 0.155~{\mathrm{fm}^{-3}}$ for LS220 EoS. The DU process thresholds of $npe$ and $npe\mu$ matter correspond to the dotted and solid black curves, respectively. As in Fig. 8, the symmetry energies at the central baryon number density are plotted for $1.1, 1.4, 1.8,$ and $2.1~M_{\odot}$ models with Togashi and Shen EoSs, and  $1.1, 1.4, 1.7$ and $2.0~M_{\odot}$ models with LS220 EoS.}
\label{fig:su}
\end{figure}

As described in the previous section, the EoS determines whether the DU process turns on or not. This is because $Y_p$ depends on the EoS. In our calculations, EoSs are assumed to be in the $\beta$ equilibrium and charge neutrality:
\begin{equation}
\mu_n - \mu_p = \mu_e = \mu_\mu,
\label{eq:betaeq}
\end{equation}
\begin{equation}
Y_p = Y_e + Y_\mu,
\label{eq:chrgnt}
\end{equation}
where $\mu_n$, $\mu_p$, $\mu_e$, and $\mu_{\mu}$ are the chemical potentials of neutrons, protons, electrons, and muons, respectively. In Fig.~\ref{fig:yp}, we show $Y_p$ as a function of the baryon mass density for the EoSs adopted in this study. In this figure, adopting the LS220 EoS, we also plot the DU threshold, $Y^\mathrm{DU}_p$, obtained by substituting $Y_e$ and $Y_\mu$ in Eq.~(\ref{eq:7}). The DU process turns on for $Y_p \ge Y^\mathrm{DU}_p$. We can recognize that, with use of the LS220 EoS, the DU process turns on in neutron stars with $\gtrsim$1.4$M_\odot$ but this is not the case for neutron stars with $1.1M_\odot$. In case of the Togashi EoS, $Y_p$ is lower than the DU threshold and, therefore, the DU process is disallowed even for the maximum-mass neutron stars. In contrast, since $Y_p$ is higher than the DU threshold, the DU process turns on for all the models of the Shen EoS considered in our calculations. Note that, whereas $Y_p^{\mathrm{DU}}$ depends on the EoS, the difference of $Y_p^{\mathrm{DU}}$ among the EoS models is insignificant compared to that of $Y_p$. 

For the EoS of neutron star matter, $Y_p$ is related to the symmetry energy, $S(u)$. When the symmetry energy is defined as the energy difference between the pure neutron matter and the symmetric nuclear matter as in Eq.~(\ref{eq:9}), the chemical potentials of neutrons and protons satisfy the relation:
\begin{equation}
\mu_n - \mu_p = 4S(u)(1 - 2Y_p).
\label{eq:chemsym}
\end{equation}
Then, $Y_p$ is determined by Eqs.~(\ref{eq:betaeq}), (\ref{eq:chrgnt}) and (\ref{eq:chemsym}). In general, $Y_p$ is smaller for the EoS with lower symmetry energy, provided that the baryon number density is the same. This trend is confirmed for the EoSs adopted in this study from Fig.~\ref{fig:su}, where the symmetry energy is shown as a function of the baryon number density. At high densities, the Togashi EoS has the lowest symmetry energy while the Shen EoS has the highest. This is consistent with the fact that the symmetry energy slope parameter, $L$, is small for the Togashi EoS but is large for the Shen EoS. Whereas the value of $L$ is defined at the saturation density, it can be used as an indicator of the symmetry energy at high densities.

As a result of the discussion presented so far, the DU process is disallowed even for heavy neutron stars if the EoS adopted for the cooling calculations has low symmetry energy at high densities, or, small $L$ value. Here, we consider the critical symmetry energy as a function of density, $S^{\mathrm{DU}}(u)$, for the DU process. When muons are absent, the charge neutrality requires $Y_e=Y_p$ and, therefore, $S(u)$ and $Y_p$ satisfy the relation \cite{Lattimer1991a,Tews2017}:
\begin{equation}
\mu_n - \mu_p = 4S(u)(1 - 2Y_p) = \frac{hc}{2\pi}(3\pi^2n_0 uY_p)^{1/3} = \mu_e,
\label{eq:nomuon}
\end{equation}
with the Planck constant $h$ and the nuclear saturation number density $n_0$. Then, the condition for the DU process is $Y_p\ge Y^{\mathrm{DU}}_p=1/9$, or,
\begin{equation}
S(u) \ge S^{\mathrm{DU}}(u) \simeq 50.7 \left( \frac{un_0}{0.155~\mathrm{fm}^{-3}} \right)^{1/3}~\mathrm{MeV}.
\label{eq:nomuons}
\end{equation}
Similarly, taking into account the muon contributions, we can obtain $Y_p$, $Y_e$, and $Y_\mu$ from Eqs.~(\ref{eq:betaeq}), (\ref{eq:chrgnt}) and (\ref{eq:chemsym}) if the symmetry energy $S(u)$ is provided as a function of the density $u$. Then, we can evaluate the DU threshold $Y^{\mathrm{DU}}_p$ by Eq.~(\ref{eq:7}) and judge whether the DU process turns on. The critical symmetry energy $S^{\mathrm{DU}}(u)$ derived in this way is plotted in Fig.~\ref{fig:su}. As seen in this figure, the difference of $S^{\mathrm{DU}}(u)$ between $npe$ and $npe\mu$ matter is a few MeV.

Since the DU process is disallowed for any masses, the models with the Togashi EoS are inconsistent with the observations of low-temperature INS as described in the previous section. It stems from the fact that the Togashi EoS has low symmetry energy at high densities, or, small $L$ value. In contrast, a small $L$ value is preferred from the observations of neutron star radius, which indicates the small value for the radius. Therefore, we emphasize that we should consider not only the radius but also the cooling curve so as to examine the neutron star EoS. Note that the fast cooling processes due to exotic particles are omitted in the above discussion.

\subsection{Influence of the surface composition on cooling curves}

\begin{figure}[t]
\centering
\includegraphics[width=5.0in]{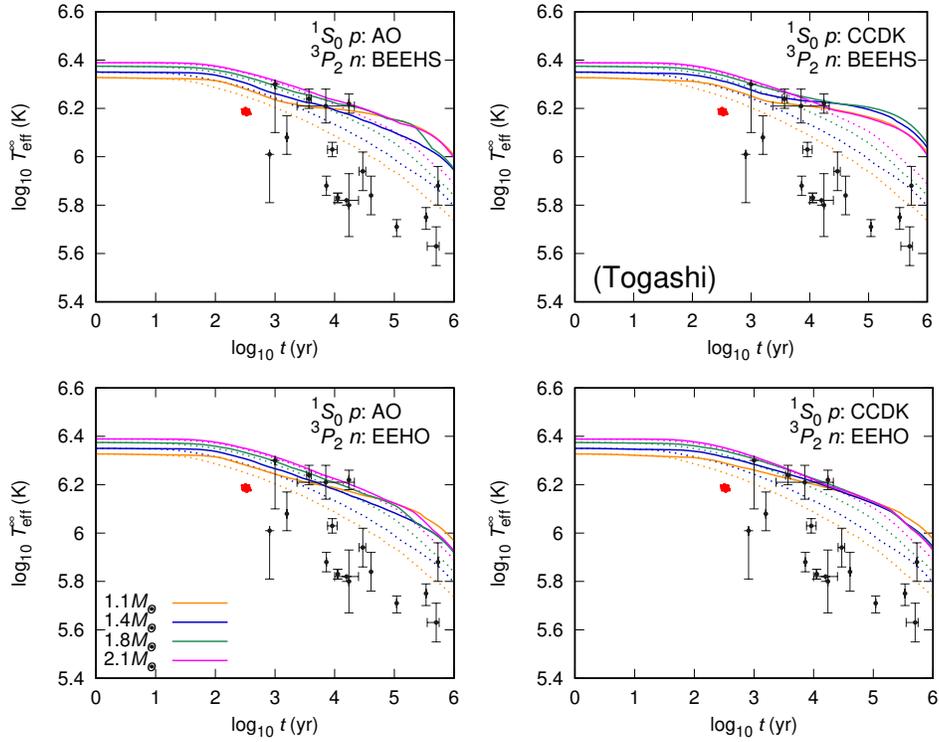}
\caption{Same as Fig. 4 but with pure-Ni surface}
\end{figure}

In our cooling calculations as described in the previous section, the neutron star surface is assumed to have the light compositions, which are 73\% of ${}^{1}{\mathrm{H}}$, 25\% of ${}^{4}{\mathrm{He}}$, and 2\% of ${}^{56}{\mathrm{Ni}}$. Many previous studies show that the surface compositions affect the effective surface temperature of INS \cite{Noda2006,Page2006}. Therefore, using the Togashi EoS, we calculate cooling curves with the pure-Ni surface, which is the heaviest element in conceivable compositions. The results are shown in Fig. 10. Compared to Fig. 4, the overall effective temperature of INS with pure-Ni surface is lower than that of INS with light compositions. In addition, when the age of INS becomes $10^{5-6}~\mathrm{yr}$ at the photon cooling stage, INS with light surface compositions cool rapidly. We see however that these modifications of surface compositions on cooling curves are insufficient to account for the observed data of INS with low temperature. This is similar to the result of Ref. \cite{Beznogov2016} (see Figure 4 in it). Therefore, the cooling curves with use of the Togashi EoS are inconsistent with the INS with low surface temperature, even if the INS surface is composed of heavy elements.

\section{Concluding remark}

We have performed cooling simulation of neutron stars with use of the Togashi, Shen and LS220 EoSs. As a result, the LS220 EoS can account for the temperature observations of INS through the DU process by considering appropriate superfluid model. On the other hand, the Togashi EoS is inappropriate to account for the observations of INS regardless of surface composition because the DU process is prohibited. The fundamental reason comes from rather low symmetry energy, which is appropriate for the neutron star radius. Therefore, other fast cooling processes than the DU process may work in INS with low temperature. For example, the neutrino emissions concerned with exotic particles such as hyperons, pions, kaons, and quarks lead to rapid cooling (e.g. \cite{Lim2019,Raduta2019}). Since the value of the symmetry energy is predicted to be small and the DU process is hard to occur, we can imply the possibility that some exotic particles exist in neutron stars. We will investigate their impacts on the cooling curve, as well as on the EoS, elsewhere. Furthermore, the critical temperature of the superfluid transition also depends on the EoS. Calculations taking into account the consistency between the EoS and neutrino emissivity are worth performing.

%
%
%
%

\end{document}